\documentclass[]{aa}
\usepackage{psfig}
\begin{document}

\thesaurus{06(08(08.02.1, 08.09.2))}

\title{Spectrophotometry and period analysis of the sdB eclipsing
binary HW~Virginis}

\author{L.L. Kiss\inst{1,3} \and B. Cs\'ak\inst{2} \and K. Szatm\'ary\inst{1}
\and G. F\H{u}r\'esz\inst{1,3} \and K. Szil\'adi\inst{1,3}}

\institute{Department of Experimental Physics and Astronomical Observatory,
University of Szeged,
Szeged, D\'om t\'er 9., H-6720 Hungary \and
Department of Optics \& Quantum Electronics, University of Szeged,
POB 406, H-6701 Szeged, Hungary \and
Guest Observer at Konkoly Observatory}

\titlerunning{HW Virginis}
\authorrunning{Kiss et al.}
\offprints{l.kiss@physx.u-szeged.hu}
\date{Received 11 October 2000, in original form 18 July 2000,
accepted 19 October 2000}

\maketitle
 
\begin{abstract}

New CCD observations of the pre-cataclysmic binary HW~Virginis are
presented and discussed. The R-filtered CCD photometry
was supplemented with medium-band (Cousins VRI) spectrophotometry
based on low-resolution objective-prism spectra.
The period variation is reanalysed by means of the standard O--C technique.
The new data support the conclusions of Kilkenny et al. (2000)
on the strong and continuous period change.
The long-term period variation can be described approximately
with two linear branches
in the O--C diagram corresponding to a sudden period jump between two
constant periods around JD 2448500 (1991). Additional smooth small-scale
changes in the period distort the linearity.
The present data do not support the hypothetical light-time effect
of \c Cakirli \& Devlen (1999).

The eclipse depths in V, R, and I do not
show colour dependence, which suggests negligible continuum
variations due to the cool secondary component in the far-red region
(up to 8800 \AA). The magnitude of the reflection effect was used
to estimate the mean effective temperature of the illuminated
hemisphere of the secondary. The result is 13300$\pm$200 K or
11000$\pm$200 K, depending on the primary temperature (35000 K or
26000 K). An albedo near unity is implied for the cool component.

 
\end{abstract}

\section{Introduction}

The eclipsing binary nature of HW~Vir (BD$-07^\circ$3477,
V$_{\rm max}$=10\fm5, P=0.1167 days) was discovered by Menzies (1986).
Before the discovery it was identified as
a UV-bright object (Carnochan \& Wilson 1983).
Both primary and secondary minima are observed, and 
the lightcurve exhibits a striking 
reflection effect with an amplitude about 0\fm2.
Different lightcurve solutions were published (Menzies \& Marang 1986,
Wood et al. 1993, W\l odarczyk \& Olszewski 1994, \c Cakirli \& Devlen 1999).
Although there are some discrepancies between the models presented, the
general appearence is well-defined: the primary is a bright
and evolved sdB star being overluminous compared to the
low luminosity and cool secondary star, which is most probably
an M-type main-sequence star. The temperature estimates
range between 26000--36000 K and 3200--3700 K for the primary and
secondary, respectively.
The large brightness
difference has prevented direct spectroscopic detection
of the cool component, while Wood \& Saffer (1999) reported extra
H$\alpha$ absorption features around the maximal reflection
effect, which were associated with the illuminated and heated
secondary atmosphere.
The relatively large uncertainty of the primary temperature is
reflected in the published distance estimates ranging
from 42-151 pc (Wood et al. 1993), 125 pc (W\l odarczyk \& Olszewski
1994), 210 pc (calculated from parameters in \c Cakirli \& Devlen 1999),
171$\pm$19 pc (Wood \& Saffer 1999) and 179 pc (also derived from
parameters in Hilditch et al. 1996). In contrast to these
values, direct astrometric measurements by the Hipparcos
satellite resulted in a parallax of 1.8$\pm$1.9 mas (ESA 1997), i.e.
the star {\it may} be at a much larger distance than was thought earlier.

The first note on the strong
period decrease in HW~Vir was published by Kilkenny et al. (1994).
They discussed the period change over a 9-year
baseline and pointed out possible reasons. They suggested
angular momentum loss via magnetic braking in a modest
stellar wind to be the most likely possibility for the
period decrease, though light-time effects due to
orbital motion around a third body were not ruled out.
This latter approach was revised by \c Cakirli \& Devlen (1999),
who presented a light-time effect solution
of the O$-$C diagram, though the observations covered only 69\% of
the suggested orbital period. This has been the initiator
of our photometric observations, because we wanted to check
the recent period change. Very recently, Kilkenny et al. (2000)
analysed an updated O$-$C diagram concluding that even a
6th order polynomial fit is not perfect.

The main aim of this paper is to present an analysis of
new photometric and low-resolution spectroscopic observations
of HW~Vir carried out in May, 2000. The paper is
organised as follows: Sect.\ 2 deals with the observations
and data reductions. The period analysis is presented in Sect.\ 3,
while several simple considerations based on the multicolour
spectrophotometry are given in Sect.\ 4. A brief summary
of the presented results is listed in Sect.\ 5.

\section{Observations and data reductions}

HW~Vir was observed on eight nights at two observatories in May 2000.
CCD photometry was carried out on four nights at the University of Szeged
using a 0.28-m Schmidt-Cassegrain telescope (SCT) located in
the very center of the city of Szeged. The detector was
an SBIG ST-6 CCD camera (375x242 pixels) giving an angular
resolution of about 2 "/pixel (the pixels are rectangular).
The observations were mostly through an R filter.

As the primary minimum lasts for only
20 minutes, we chose 30-second exposures. This way we could
avoid the phase smearing of the lightcurve.
Two comparison stars were chosen in the field
(comp = GSC 5528-0591, 12\fm4, check = GSC 5528-1273,
12\fm3). Throughout the observations we did not find significant
variations of their brightness difference.

The CCD frames were reduced with standard tasks in IRAF.
The dark-corrected frames were flat-fielded with an average sky-flat
image obtained during the morning twillight after every night.
We did aperture photometry with IRAF/DAOPHOT because the large pixels
prevented psf-photometry; the field is uncrowded in any case.
The photometric accuracy was estimated from the scatter in
the difference between the comparison stars. This yielded an uncertainty of
$\pm$0\fm03 in R-band.

We have determined five new times of minimum (both primary and
secondary) by fitting low-order (3-5) polynomials to
the observed lightcurve points centered on the primary and
secondary minima.
The differential lightcurves were phased according to the
new ephemeris (see later). The resulting phase diagram based
on 299 individual points is shown in Fig.\ 1.

\begin{figure}
\begin{center}
\leavevmode
\psfig{figure=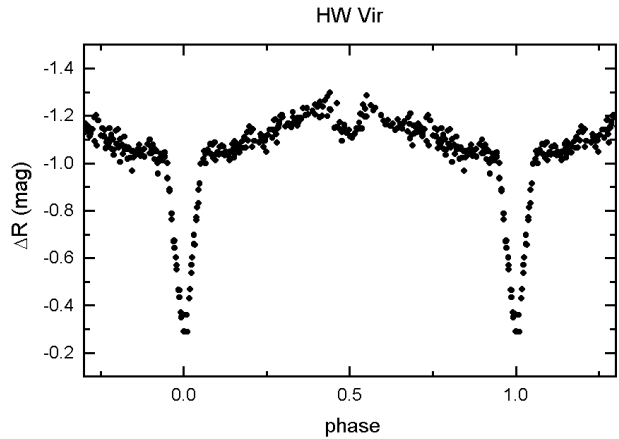,width=\linewidth}
\caption{The R-band lightcurve phased with the new ephemeris
(E$_{\rm 0}$=2451670.4125, P=0.116719411 days).}
\end{center}
\label{f1}
\end{figure}

\begin{table}
\begin{center}
\caption{Journal of observations}
\begin{tabular} {llll}
\hline
Date         &  Type        & Instr. &  Length\\
2000 ..      &              &        &        \\
\hline
May 3/4 & photometry (R)  & 0.28-m SCT & 1$^{\rm h}$\\
May 5/6 & photometry (R)  & 0.28-m SCT & 5$^{\rm h}$\\
May 6/7 & photometry (R)  & 0.28-m SCT & 5$^{\rm h}$\\
May 7/8 & photometry (V)  & 0.28-m SCT & 2$^{\rm h}$\\
May 25/26 & low-res. spec. & 0.6-m Schmidt & 3$^{\rm h}$\\
May 26/27 & low-res. spec. & 0.6-m Schmidt & 3$^{\rm h}$\\
May 27/28 & low-res. spec. & 0.6-m Schmidt & 1$^{\rm h}$\\
May 30/31 & low-res. spec. & 0.6-m Schmidt & 1$^{\rm h}$\\
\hline
\end{tabular}
\end{center}
\end{table}

We obtained digital objective-prism spectra on four nights
at the Piszk\'estet\H{o} Station of Konkoly Observatory with
the 60/90/180 cm Schmidt-telescope. The detector was a
Photometrics AT200 CCD camera (1536x1024 pixels, KAF-1600 chip
with UV-coating). The objective prism has a refracting angle
of 5$^\circ$ giving 580 \AA/mm resolution at H$\gamma$.
The observations were unfiltered in order
to detect the entire spectral region (approximately between
3800 \AA\ and 9000 \AA) limited only by the spectral
response of the CCD and the atmospheric transmission.
We took 2- to 5-minute exposures, depending on the target brightness
and weather conditions. The full log of observations
is summarized in Table\ 1.

\begin{figure}
\begin{center}
\leavevmode
\psfig{figure=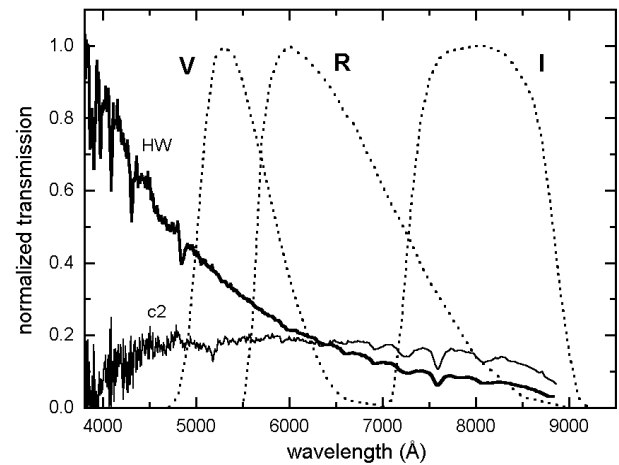,width=\linewidth}
\caption{Sample spectra of HW~Vir and GSC 5528-0580 (c2)
with the VRI passband functions taken from Bessell (1990). Note,
that the spectra are scaled arbitrarily and independently for
clarity.}
\end{center}
\label{f2}
\end{figure}

The spectral extraction was done with routines developed by the authors.
We followed the basic strategy of
automatic spectrum detection outlined by Bratsolis et al. (1998).
It consists of two main points: i) peak-finding in an integrated
profile of the image perpendicular to the dispersion; ii)
detection of the end-points of each spectrum.
For the wavelength calibration of the extracted spectra, we
used two comparison-star spectra with unambiguous spectral features.
The first one was Vega (spectral type A0), where the prominent hydrogen
Balmer-series lines provided good calibrator data (from H$\alpha$ to H7),
while the second one was $\delta$ Vir
(spectral type M3). The common features (strong atmospheric telluric lines
at 7200 \AA, 7600 \AA, 8170 \AA\, identified using the low-resolution
spectral atlas of Torres-Dodgen \& Weaver 1993), helped adjust the
spectra to the same wavelength scale.
The eleven well-defined
spectral features gave a dispersion curve between 3800 and 8800 \AA\ (with
a slight extrapolation at the red end).
The residual scatter of calibrating spectral lines is typically
about 1 pixel, corresponding to 10--20 \AA\ depending on the spectral region.
Finally, we made a relative flux-calibration by dividing the
extracted spectra with spectral response function of the
instrument (this is a product of the atmospheric
transmission and spectral sensitivity function of the CCD chip).
It was calibrated with the same spectrum of Vega, knowing its
tabulated absolute flux spectrum taken from Gray (1992).
Although we observed Vega at the same airmass as HW~Vir, a
few (2-3) percent uncertainty of the flux levels cannot
be excluded due to the possibly changing atmospheric
transmission (especially at the blue end of the spectra).

The wavelength and flux-calibrated spectra of HW~Vir and two
bright stars in the field were used to calculate standard
photometric magnitude differences.
The used comparisons were c1=GSC 5528-0591, 12\fm4
and c2=GSC 5528-0580, 12\fm4. We plot sample spectra for
HW~Vir and GSC~5528-0580 in Fig.\ 2, where standard passbands
of V, R, and I are also shown (Bessell 1990).
By multiplying the calibrated spectra with the filter transmission
functions and integrating over wavelength, we determined standard
differential V, R, and I magnitudes. Unfortunately, both comparison
stars are faint in the blue region, where HW~Vir is considerably
brighter (see the large scatter below 4500 \AA\ in the spectrum of
c2 in Fig.\ 2), that is why we chose only the VRI bands. The phase
diagrams from the obtained
157 points are shown in Fig.\ 6. The c1$-$c2 differences implied
an estimated photometric accuracy of $\pm$0\fm015 in V
and $\pm$0\fm01 in R and I bands.
 Times of primary and secondary
minima were determined with the same method as applied in CCD
photometry, however, these epochs are of lower accuracy due to
the meagre phase coverage caused by the relatively long exposures
(2-5 min). The eclipse depths are also affected by this
phase smearing. All data presented in this paper are available upon
request from the first author.

\section{Period analysis}

The period variation of HW~Vir was studied by means of the standard
O$-$C technique. For this,
we have collected all available times of primary and secondary
minimum.
\c Cakirli \& Devlen (1999) gave a nearly complete compilation of
times of minimum between 1984 and 1997, which had to be updated
mainly with the newly published data. These are listed in Table\ 2.
We excluded all of the secondary minima from our analysis,
because the depth and sharpness of the primary
minimum make the timing more accurate. This was obvious when
plotting the whole O$-$C diagram,
where the most discordant points were of secondary minima.
To be consistent we omitted all secondary minima
from the analysis, though some of them were in good agreement with
the primary ones. The final sample contains 144 points.

\begin{table*}
\begin{center}
\caption{Times of minima of HW~Vir (HJD$-$2400000). References: (1) ESA 1997,
(2) Agerer \& H\"ubscher 1996, (3) Selam et al. 1999, (4) Agerer et al. 1999,
(5) Kilkenny et al. 2000, (6) Ogloza et al. 2000,
(7) this paper, CCD-R photometry, (8) this paper, objective-prism
photometry}
\begin{tabular} {|lll|lll|lll|lll|}
\hline
Min  & Type & Ref. & Min  & Type & Ref. & Min & Type & Ref. & Min  & Type & Ref.\\
\hline
48500.0801    & pri. & 1$^{\rm a}$ & 50155.5119    & pri. & 3 & 50552.4744    & pri. & 6 &   50948.3866    & pri. & 4\\                                                        
49190.24208   & pri. & 5 &           50185.39198   & pri. & 5 & 50575.46820   & pri  & 5 &   50955.38977   & pri  & 5\\                                                        
49418.54535   & pri. & 5 &           50186.44247   & pri  & 5 & 50594.3768    & pri. & 6 &   50959.24150   & pri. & 5\\                                                        
49437.57069   & pri. & 5 &           50201.38254   & pri. & 5 & 50597.29471   & pri. & 5 &   51021.21952   & pri. & 5\\                                                        
49450.64320   & pri. & 5 &           50202.43302   & pri. & 5 & 50599.27895   & pri  & 5 &   51183.57618   & pri  & 5\\                                                        
49476.32145   & pri  & 5 &           50216.67281   & pri  & 5 & 50600.32943   & pri. & 5 &   51190.57932   & pri. & 5\\                                                        
49480.40663   & pri. & 5 &           50218.42364   & pri. & 5 & 50604.4147    & pri. & 6 &   51216.49105   & pri. & 5\\                                                        
49485.30883   & pri. & 5 &           50222.50883   & pri. & 5 & 50631.26003   & pri. & 5 &  51236.56678   & pri  & 5\\                                                         
49496.397     & pri. & 2 &           50280.28481   & pri  & 5 & 50883.49067   & pri  & 5 &  51300.4125    & pri. & 6\\                                                         
49519.27421   & pri  & 5 &           50491.4300    & pri. & 3 & 50885.47490   & pri. & 5 &  51301.3460    & pri. & 6\\                                                         
49728.55216   & pri. & 5 &           50491.4886    & sec. & 3 & 50910.45284   & pri. & 5 &  51301.4629    & pri. & 6\\                                                         
49733.57107   & pri. & 5 &           50491.5467    & pri. & 3 & 50912.55379   & pri. & 5 &  51668.4288    & pri. & 7\\                                                         
49778.6249    & pri. & 6 &           50506.48700   & pri. & 5 & 50913.50427   & pri. & 5 &  51670.3550    & sec. & 7\\                                                         
49785.6279    & pri. & 6 &           50509.52176   & pri. & 5 &   50927.3774    & pri. & 4 & 51670.4125    & pri. & 7\\                                                        
49808.5048    & pri. & 6 &           50510.57223   & pri  & 5 &   50927.4938    & pri. & 6 & 51670.4717    & sec. & 7\\                                                      
49833.48294   & pri  & 5 &           50511.5057    & pri. & 3 &   50931.34559   & pri. & 5 & 51671.4632    & pri. & 7\\                                                      
49880.28740   & pri. & 5 &           50511.50598   & pri. & 5 &   50943.3678    & pri. & 4 & 51691.3643    & sec. & 8\\                                                       
50142.55596   & pri. & 5 &           50511.5636    & sec. & 3 &   50943.4262    & pri. & 4 & 51691.4225    & pri. & 8\\                                                       
50144.54015   & pri  & 5 &           50543.72054   & pri  & 5 &   50943.4843    & pri. & 4 & 51692.3561    & pri. & 8\\                                                      
50147.57490   & pri. & 5 &           50547.45557   & pri. & 5 &   50946.4024    & pri. & 4 & 51695.3905    & pri. & 8\\                                                      
50155.3946    & pri. & 3 &                         &      &   &                 &      &   &               &      &  \\
\hline                                                        
\end{tabular}
\end{center}
$^{\rm a}$ The Hipparcos Epoch Photometry database
gives an incorrect doubled period for HW~Vir; the listed
epoch was determined by us.
\end{table*}

\begin{figure}
\begin{center}
\leavevmode
\psfig{figure=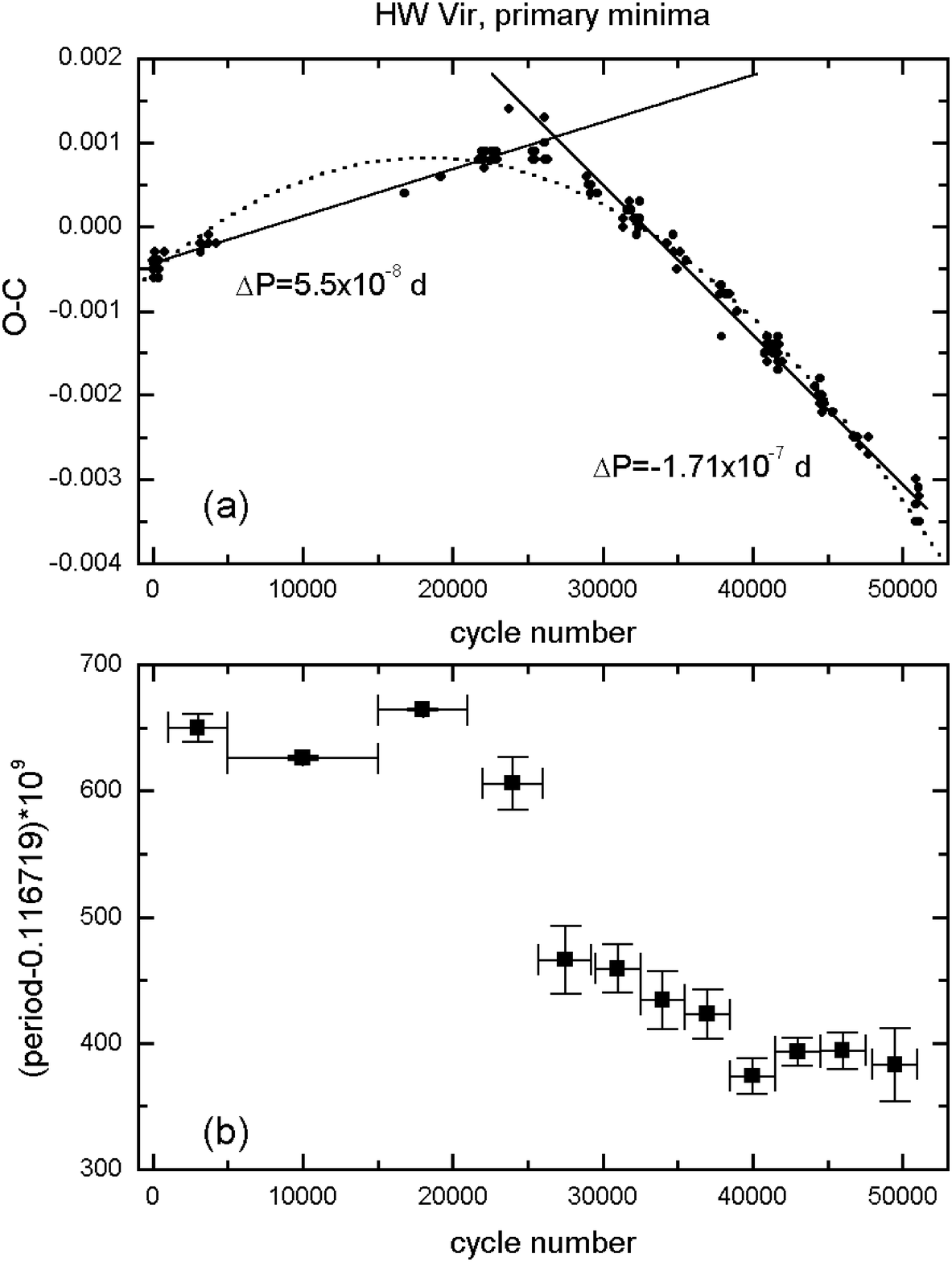,width=\linewidth}
\caption{{\bf (a):} The O$-$C diagram of HW~Vir.
{\bf (b):} The ``instantaneous period'' of HW~Vir. The horizontal
error bars mark the distance of the neighbouring seasons, while
the vertical ones are the standard errors of the linear fits. }
\end{center}
\label{f3}
\end{figure}

We calculated the O$-$C diagram with the following elements taken
from \c Cakirli \& Devlen (1999):

\begin{center}
Hel. JD$_{\rm min}=2445730.5565+0.\!\!^{\rm d}1167195820 \cdot E$
\end{center}

\noindent It is plotted in the top panel of Fig.\ 3.
Our new O$-$C diagram covering almost 52000 cycles (17 years)
does not support either periodic solutions with light-time effect
or continuous period decrease via smooth and slow mechanisms.
One of the simplest and statistically most likely descriptions is
assuming two linear branches with a sudden change around
cycle number 25000. We fitted two separate models and found
a period difference of $(2.25\pm0.04) \cdot 10^{\rm -7}$ d,
a slightly larger value than the $(1.86\pm0.09) \cdot 10^{\rm -7}$ d
determined by Wood \& Saffer (1999).
We conclude that before 1991
the period was 0$.\!\!^{\rm d}$116719636(1), which changed
to 0$.\!\!^{\rm d}$116719411(4).

We have examined the possibility of continuous
period decrease. For this reason we fitted a parabola
to the O$-$C diagram shown as dotted line in the top
panel of Fig.\ 3. The resulting standard deviation of the residuals
is somewhat larger than in the case of two linear
fits ($2.1\cdot10^{\rm -4}$ vs. $1.9\cdot10^{\rm -4}$).
Nevertheless, one could argue that the individual deviations
of the O$-$C points from the fitted parabola may be due
to some kind of systematic effect, e.g. light-time
effect in an undetected binary system. This is one of the
usual ways of interpreting cyclic or quasi-cyclic O$-$C diagrams of
eclipsing binaries (see, e.g., Borkovits \& Heged\"us 1996),
thus we have tried to explain the secular period change following
this approach. After the parabola subtraction the residuals have
a cyclical behaviour, which is the most important condition
for assuming a periodic light-time effect solution.
However, this possibility has been ruled out with a simple
numerical test performed as follows. An artificial
O$-$C diagram was calculated with two linear elements
divided at the middle of the data (cycle numbers were
adjusted to the real case of HW~Vir). The O$-$C values
were truncated below the expected order of magnitude of accuracy
of photoelectric times of minima (0.0001 days).
Then we fitted a parabola to the artificial data and
the residuals after the subtraction were remarkably similar
to the real data. Thus we had
to reject the hypothetical light-time effect.

We have also determined the ``instantaneous'' period from the
neighbouring seasons. Its variation can be seen in bottom panel
of Fig.\ 3.
We conclude that there are smooth, small-scale, but not strictly
repeating changes of the period besides a large jump that happened
in less than a year around 1991. These conclusions are substantially
very similar to those of Kilkenny et al. (2000), only the time basis
is longer by a year.

Accepting two linear parts of the O$-$C diagram, we
determined the actual ephemeris, which is very
important for effective planning of follow-up observations.
It can be used to predict orbital phases provided
the period does not change again:

\begin{center}
Hel. JD$_{\rm min}=2451670.4125+0.\!\!^{\rm d}116719411(4) \cdot E$
\end{center}

\noindent These elements were used to construct every phase diagram
throughout the paper.

What is the reason for the sudden period change? The most common
assumption is mass transfer, however in a detached system such as HW~Vir
it is not expected. Nevertheless, adopting mass transfer to be
the responsible cause for period decrease, the primary star
had to transfer a mass of about $\sim3.5\cdot10^{\rm -5}$ M$_\odot$.
But, as was noted by Wood \& Saffer (1999), it is unclear
why such an event would occur.
One could consider outer accretion, e.g., engulfing
a planet-size body. The required mass (double the amount above to get
the same change of the mass-ratio) is only in the range
of tens of Earth masses, however, since we have no
additional data, this is only rough speculation.

\section{The components of HW~Vir}

\begin{figure}
\begin{center}
\leavevmode
\psfig{figure=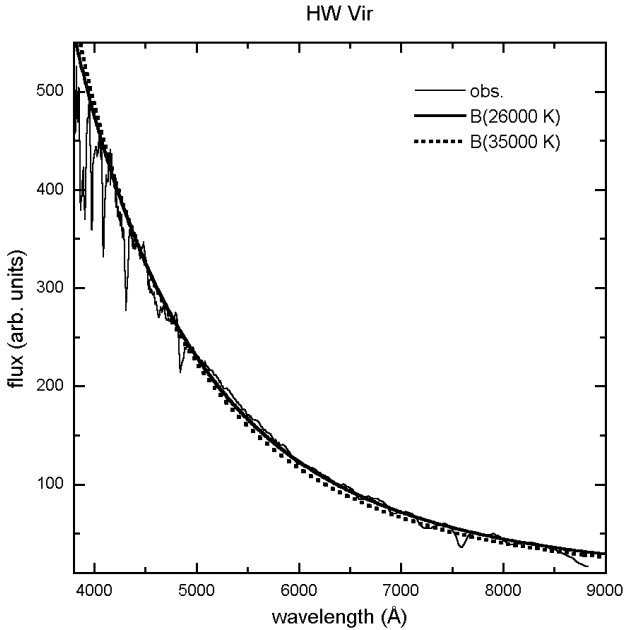,width=\linewidth}
\caption{The observed spectrum at $\phi=0.93$ with two
blackbody fits.}
\end{center}
\label{f4}
\end{figure}

\begin{figure}
\begin{center}
\leavevmode
\psfig{figure=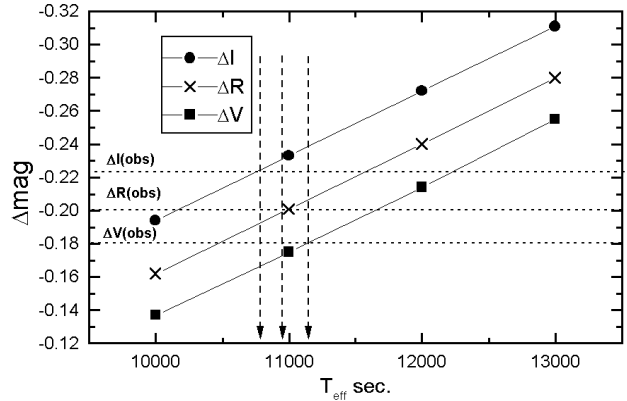,width=\linewidth}
\caption{A multiband estimate for the disk-averaged temperature of the
illuminated surface of the secondary component. The ``obs'' subscript
and the dotted lines correspond to the observed amount of the reflection
effect (i.e. the magnitude difference of the V, R, and I band
lightcurves between $\phi=0.58$ and $\phi=0.93$). The model lines
were calculated with blackbody approximation (T$_{\rm eff}$(pri.)=26000 K).}
\end{center}
\label{f5}
\end{figure}

\begin{figure}
\begin{center}
\leavevmode
\psfig{figure=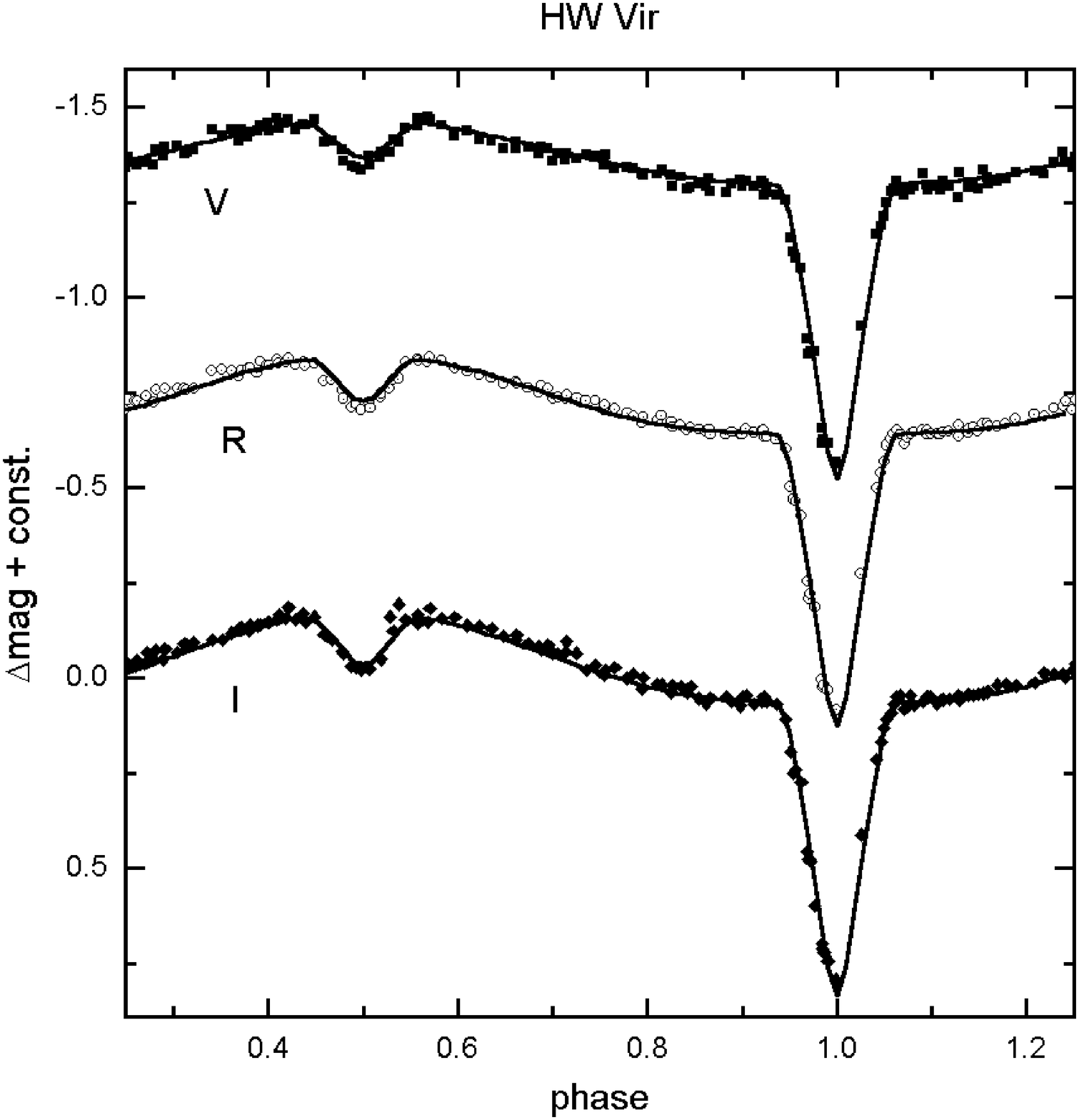,width=\linewidth}
\caption{Lightcurve models for HW~Vir}
\end{center}
\label{f6}
\end{figure}

Beside revealing the current period variation of HW~Vir, our observations
addressed the spectroscopic detection of the cool
secondary component. One possible detection of this
faint star was presented by Wood \& Saffer (1999), who found
evidence for weak additional H$\alpha$ absorption lines around
the maximal reflection effect (at $\phi$=0.58). They attributed
these lines to irradiation of the face of the secondary star
closest to the sdB star. However, there has been no identification
of the secondary during the primary minimum, when the system is
seen from the direction of the cooler hemisphere.
Our spectral coverage is only slightly extended toward the
infrared than that of Wood \& Saffer (1999), which has been
the ``reddest'' spectroscopy so far (up to 8667 \AA).
Since the eclipse depth of the primary minimum is the
same in all three bands (0\fm72), we conclude that
we have not detected the red continuum of the secondary.

The most striking feature of the lightcurve is the
prominent reflection effect. As has been noted by
W\l odarczyk \& Olszewski (1994), the effect gets
stronger with the increasing wavelength (they
obtained 0\fm17 in B, 0\fm18 in V and
0\fm21 in R).
We have also determined the amount of reflection as the
magnitude difference between $\phi$=0.58 and $\phi$=0.93.
We obtained 0\fm18 in V, 0\fm20 in R and
0\fm22 in I the
first two values being in good agreement with those of
W\l odarczyk \& Olszewski (1994). They were used
to estimate the mean heating of the secondary.

In what follows we outline a simple consideration on the
reflection effect based on blackbody approximation. We have
to stress that we did not want to present a new Wilson--Devinney-type
lightcurve solution, but we use their constraints
on the components.
All of the published lightcurve models are in
good agreement concerning the radii for
both components, approximately the same value around 0.2 R$_\odot$.
The presented values range from 0.18 R$_\odot$ to 0.22 R$_\odot$, thus
a $\sim$10\% uncertainty can be roughly assumed. Fortunately,
all models in the literature agree in one aspect: both stars
have essentially the same radius, which is the most important
assumption in what follows.
The most significant difference is related with the
temperature of the primary: the estimates are between
26000 K (W\l odarczyk \& Olszewsi, 1994) and 36000 K
(\c Cakirli \& Devlen 1999). To obtain an independent
estimate we fitted the continuum observed
just before the primary minimum ($\phi=0.93$) with a Planck function.
In the chosen phase the illuminated secondary hemisphere
is hidden and the primary totally outshines the secondary.
This can be seen in Fig.\ 4, where two Planck functions
are shown with the observed spectrum. Obviously, either
26000 K or 35000 K fits almost equally well the observed
spectrum. Keeping in mind the possible few percent uncertainty
of the flux calibration, we can claim that our data give only
limits on the primary's temperature. We can solidly exlude
25000 K and 40000 K for the lower and upper limits,
respectively (the hotter limit is less determined, as suggested
from the lightcurve modelling discussed below).

We have synthesized combined spectra of primary+irradiated
secondary by co-adding two different blackbody flux distributions
with the same weighting factors (i.e. we assumed equal surfaces of the
components). First we fixed the primary's temperature at 26000 K and
varied the second blackbody's temperature between 10000 and 13000 K.
Then we calculated the standard V, R, and I magnitude
differences between the single blackbody and combined
flux distribution. We plot these magnitude differences versus
the secondary temperatures in Fig.\ 5.
The observed strengths of the reflection effect
result in a disk-averaged temperature of 11000$\pm$200 K
for the brighter hemisphere of the secondary. Repeating
this procedure with a fixed primary temperature of
35000 K we got 13300$\pm$200 K for the secondary.
These values give a constraint on the albedo of the secondary.
Accepting approximate geometric parameters of the system
(two stars with radii of 0.2 R$_\odot$ at a distance about
0.9 R$_\odot$), a perfect blackbody and flat secondary would be heated
to 14500 K (12500 K) by a 35000 K (26000 K) hot primary.
The fact that a limb-darkened average radiation of
the irradiated secondary can be approximated
by a blackbody close to the ideal case, suggests an albedo near unity.

We have calculated
synthetic binary lightcurves using the BINARY MAKER package
by Bradstreet (1993). Let us emphasize again that
we did not want to present a new lightcurve solution,
because our data are not well suited for this purpose.
For instance, the eclipse depth of the primary minimum
is decreased by 0\fm02--0\fm03 to
0\fm72 due to the relatively
long exposures applied during the faintest state (4-5 minutes),
while photometric observations with higher time-resolution
yield 0\fm75. Therefore, our
model should be considered as an approximate one.
We adopted the following parameters of the components
for calculating the model lightcurves plotted in Fig.\ 6:
q=0.3; i=80$.\!\!^\circ$2; T$_{\rm 1}$=35000 K;
T$_{\rm 2}$=3250 K (assumed); r$_{\rm 1}$=0.2; r$_{\rm 2}$=0.2;
gravity darkening 1 and 0.32; limb-darkening (1) 0.3, 0.25, 0.20 (VRI);
limb-darkening (2) 0.0, 1.0, 0.8 (VRI);
albedo (1) 1, 1, 1 (VRI); albedo (2) 0.7, 0.7, 0.8 (VRI).
We largely followed Kilkenny et al. (1998) in choosing various
parameters, as they used the same package. The assumed
binary parameters are the same as in the previous studies.
It is interesting that we did not have to choose unity albedo,
but slightly smaller values giving acceptable curves.
Generally, the models are in good agreement with the observed
lightcurve shapes, only the calculated secondary minima are
shallower by about 0\fm01 than the observed ones. They can be made
deeper by increasing the primary temperature up to 40000-44000 K, but
this is in contradiction with the wide-band flux distribution.

The temperature ambiguity recalls the question of distance to
HW~Vir. The Hipparcos parallax is 1.8$\pm$1.9 mas (ESA 1997),
where the large uncertainty is most probably due to the
faintness of the star. Furthermore, the Hipparcos parallax error
on HW~Vir is only 1-$\sigma$, thus assuming normal distribution
it means that there is 16\% chance for the parallax being larger
than 3.7 mas and 2.5\% for 5.6 mas. Briefly, the Hipparcos parallax
is not significant, essentially
unknown. The 95\% significance limit would imply a highest parallax limit
of 1.8 mas+2$\cdot \sigma=$5.6 mas corresponding to a minimal
distance of 179 pc. This lies close to the the larger range of
published photometric and/or spectroscopic distances suggesting
the close proximity (e.g. 42 pc) to be unlikely. Keeping in mind
the quoted uncertainties, the presently available trigonometric
data yield only very rough order of magnitude estimates.

\section{Summary}

The new results presented in this paper can be summarized
as follows:

\noindent 1. New photometric observations are presented
revealing the recent period change of HW~Vir.
The O$-$C diagram covering 17 years does not support earlier suggestions
of a possible light-time effect, but can be
very well described by two linear branches.
There are also additional smooth period changes.

\noindent 2. Low-resolution objective-prism spectra
were obtained and standard VRI photometry was done
by convolving the calibrated spectra with the filter
transmission functions. Our I-band lightcurve is the
first in the literature. We have tried to detect
spectral features that can be associated with the
cool secondary component (e.g. variations in the infrared
continuum), but this attempt failed. The depth of the
primary minimum is the same in all three bands suggesting
that the secondary is totally invisible even in the I band.

\noindent 3. We present a simple estimate for the mean effective
temperature of the illuminated hemisphere of the secondary
based on the V, R, and I magnitude of the reflection effect.
Their values (0\fm18, 0\fm20 and 0\fm22)
give a 13300$\pm$200 K
or 11000$\pm$200 K temperature depending on the adopted
primary temperature, which implies a near unity albedo for the
cool component. This is also supported by an approximate lightcurve
model.

\begin{acknowledgements}
This research was supported by the ``Bolyai J\'anos'' Research
Scholarship of LLK from the Hungarian Academy of Sciences,
Hungarian OTKA Grant \#T032258 and Szeged Observatory Foundation.
The warm hospitality of the staff of the Konkoly Observatory
and their provision of telescope time is gratefully acknowledged.
Fruitful discussions with J. Vink\'o improved significantly
the quality of the paper.
The authors also thank kind help and useful comments by B. Skiff.
An anonymous referee helped to make the paper clearer and conciser.
The NASA ADS Abstract Service was used to access data and references.
This research has made use of Simbad Database operated at CDS-Strasbourg,
France.
\end{acknowledgements}

\end{document}